\pdfoutput=1
\documentclass[10pt]{scrartcl}

\usepackage[utf8]{inputenc}
\usepackage{amsmath}
\usepackage{amsfonts}
\usepackage{amssymb}
\usepackage{hyperref}
\usepackage{booktabs}
\usepackage{graphicx, adjustbox}
\usepackage{caption}
\usepackage{subcaption}

\usepackage{algorithmicx}
\usepackage{algorithm}
\usepackage{algpascal}
\usepackage{algc}
\usepackage{algcompatible}
\usepackage[noend]{algpseudocode}
\algnewcommand\algorithmicto{\textbf{to}}
\algrenewtext{For}[3]%
{\algorithmicfor\ $#1 \gets #2$ \algorithmicto\ $#3$ \algorithmicdo}

\algrenewcommand{\algorithmiccomment}[2]{\hskip#1em \texttt{// #2}}

\renewenvironment{cases}{%
\left\{\begin{array}{c@{\quad : \quad}l}}%
{%
\end{array}\right.}

\DeclareMathOperator*{\argmin}{argmin}						

\renewcommand{\S}[1]{{\mathcal{#1}}}           	

\newcommand{\abs}[1]{\mathop{\left\lvert #1 \right\rvert}} 
\newcommand{\args}[1]{\mathop{\left( #1 \right)}} 

\newcommand{\norm}[1]{\mathop{\left\lVert #1 \right\rVert}}
\newcommand{\cbrace}[1]{\mathop{\left\{ #1 \right\}}}

\newcommand{\argsS}[2]{\mathop{\left( #1 \right)#2}} 

\newcommand{\normS}[2]{\mathop{\left\lVert #1 \right\rVert#2}}

\usepackage{fullpage}
\usepackage[sort,numbers]{natbib}
\newcommand{\N}{\mathbb{N}}

\newcommand{\R}{\mathbb{R}}
\DeclareMathOperator{\dtw}{dtw}
\newcommand{\commentout}[1]{}

\begin{document}

\title{\LARGE An Average-Compress Algorithm for the Sample Mean Problem under Dynamic Time Warping}
\author{\small Brijnesh Jain, Vincent Froese, and David Schultz \\[-1ex]
 \small Technische Universit\"at Berlin, Germany\\[-1ex]
 \small e-mail: brijnesh.jain@gmail.com}
 
\date{}
\maketitle

\begin{abstract}
Computing a sample mean of time series under dynamic time warping (DTW) is NP-hard. Consequently, there is an ongoing research effort to devise efficient heuristics. The majority of heuristics have been developed for the constrained sample mean problem that assumes a solution of predefined length. In contrast, research on the unconstrained sample mean problem is underdeveloped.  In this article, we propose a generic average-compress (AC) algorithm for solving the unconstrained problem. The algorithm alternates between averaging (A-step) and compression (C-step). The A-step takes an initial guess as input and returns an approximation of a sample mean. Then the C-step reduces the length of the approximate solution. The compressed approximation serves as initial guess of the A-step in the next iteration. The purpose of the C-step is to direct the algorithm to more promising solutions of shorter length. The proposed algorithm is generic in the sense that any averaging and any compression method can be used. Experimental results show that the AC algorithm substantially outperforms current state-of-the-art algorithms for time series averaging. 
\end{abstract}

\section{Introduction}

\begin{figure}[t]
\centering
\begin{subfigure}{.3\textwidth}
  \centering
  \includegraphics[width=\linewidth]{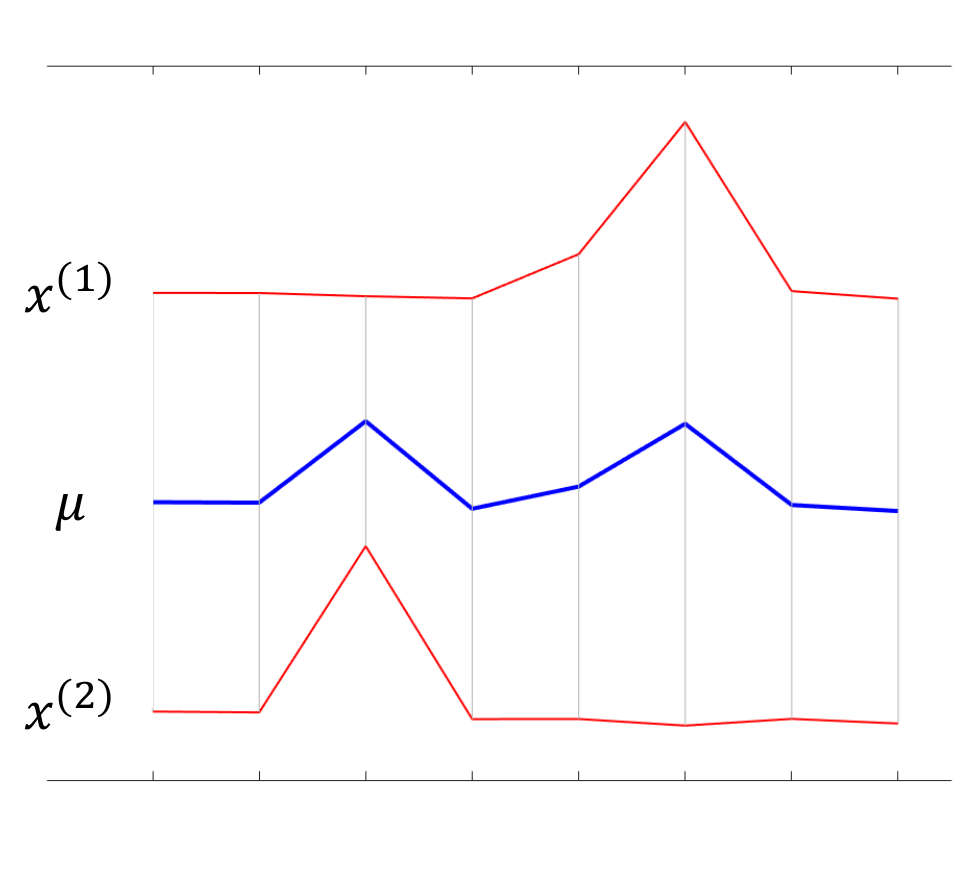}
  \caption{Arithmetic mean}
  \label{fig:euklid}
\end{subfigure}%
\hspace{1cm}
\begin{subfigure}{.3\textwidth}
  \centering
  \includegraphics[width=\textwidth]{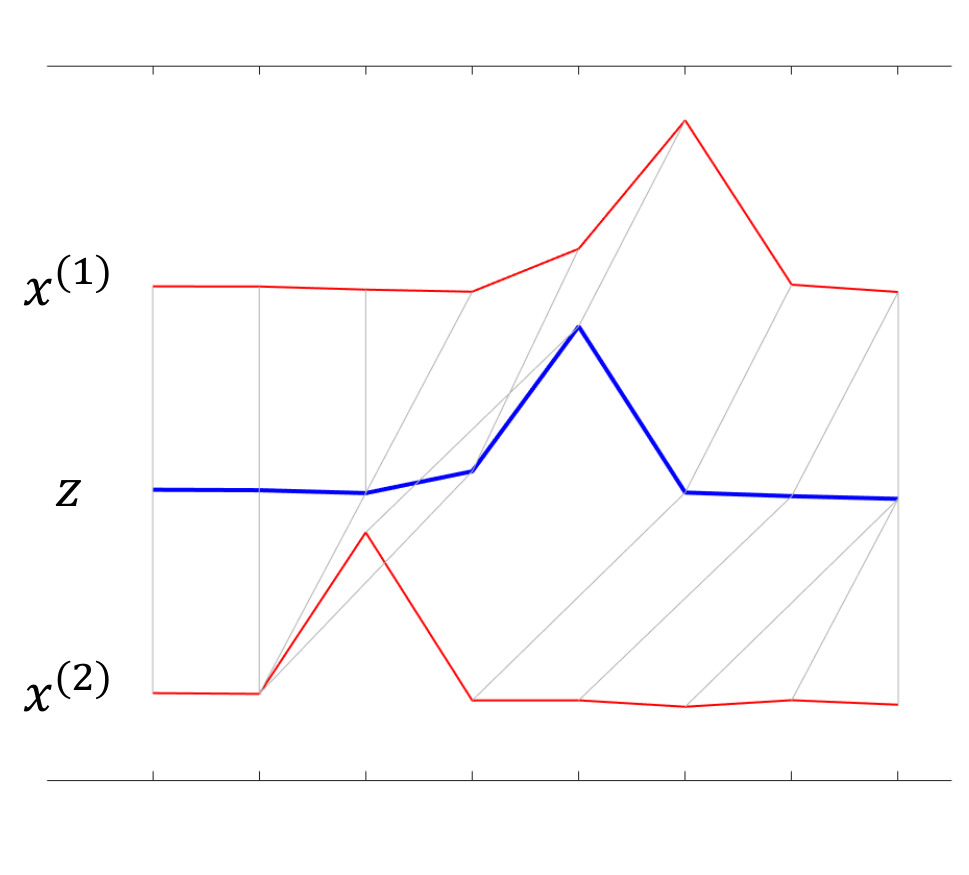}
  \caption{DTW Mean}
  \label{fig:dtw}
\end{subfigure}

\caption{Mean time series (blue) of the two sample time series $x^{(1)}$ and $x^{(2)}$ shown in red. Both time series have a single peak but are out of phase and slightly vary in speed. We may think of $x^{(1)}$ and $x^{(2)}$ as the daily average temperature of some region during the summer at two different years. Based on this information, a typical summer of this region has a single extreme heat wave. The arithmetic mean $\mu = (x^{(1)} + x^{(2)})/2$ in Fig.~\ref{fig:euklid} has two attenuated peaks suggesting that a typical summer has two moderate heat waves. In contrast, the DTW mean $z$ in  Fig.~\ref{fig:dtw} captures the characteristic properties of $x^{(1)}$ and $x^{(2)}$ and shows a single peak as a representative summary of both sample peaks.}
\label{fig:mean}
\end{figure}

Time series such as stock prices, climate data, energy usages, sales, biomedical measurements, and biometric data are sequences of time-dependent observations that often vary in temporal dynamics, that is in length, speed, and shifts in phase. For example, the same word can be uttered with different speaking speeds. Similarly, monthly temperature or precipitation extremes of certain regions can differ in duration and may occur out of phase for a period of a few weeks. 

To account for temporal variations in proximity-based time series mining, the \emph{dynamic time warping} (DTW) distance is often the preferred choice of proximity measure \citep{Aghabozorgi2015,Bagnall2017,Abanda2018}. An intricate problem in DTW-based time series mining is time series averaging. The problem consists in finding a typical representative that summarizes a sample of time series. Different forms of time series averaging have been applied  to improve nearest neighbor classifiers \cite{Jain2018,Petitjean2016,Rabiner1979}, to accelerate similarity search \cite{Tan2017}, and to formulate $k$-means clustering in DTW spaces \cite{Rabiner1979,Hautamaki2008,Petitjean2011,Soheily-Khah2016}. Figure \ref{fig:mean} presents an example illustrating why the  arithmetic mean can be inappropriate for time series averaging and motivates a concept of mean under dynamic time warping that can cope with temporal variations.

Time series averaging itself and as a subroutine of data mining tasks is inspired by the fundamental concept of mean in statistical inference. One central path in statistical inference departs from the mean, then leads via the normal distribution and the Central Limit Theorem to statistical estimation using the maximum likelihood method. The maximum likelihood method in turn is a fundamental approach that provides probabilistic interpretations to many pattern recognition methods.

This central path is well-defined in Euclidean spaces, but becomes obscure in mathematically less structured distance spaces. Since an increasing amount of non-Euclidean data is being collected and analyzed in ways that have not been realized before, statistics is undergoing an evolution \cite{Kim2010}. Examples of this evolution are contributions to statistical analysis of shapes \cite{Bhattacharya2012,Dryden1998,Kendall1984}, complex objects \cite{Marron2014}, tree-structured data \cite{Feragen2013,Wang2007}, graphs \cite{Ferrer2010,Jain2016a,Jiang2001}, and time series \cite{Brill2019,Hautamaki2008,Rabiner1979}. 

Though the volume of time series data currently collected exceeds those of the other data structures mentioned above, the concept of a mean in DTW spaces is least understood. However, a better understanding of time series averaging is the first step towards devising sound pattern recognition methods based on time series averaging such as $k$-means clustering. Examples of how the lack of a clear understanding of time series averaging may lead the field astray can be found in \cite{Brill2019,Jain2019}.

As for other non-Euclidean distance spaces,  the standard approach to time series averaging in DTW spaces is based on an idea by Fr\'echet \cite{Frechet1948}: Suppose that $\S{S} = \cbrace{x_1, \dots, x_N}$ is a sample of $N$ time series. A sample mean of $\S{S}$ is any time series $\mu$ that minimizes the Fr\'echet function 
\[
F: \S{U} \rightarrow \R, \quad z \mapsto \frac{1}{N}\sum_{i=1}^N \dtw\!\argsS{z,x_i}{^2},
\]
where $\dtw(x,y)$ is the DTW-distance and $\S{U}$ is a set of time series of finite length. The search space $\S{U}$ typically takes two forms: 
\begin{enumerate}
\itemsep0em
\item Unconstrained form: $\S{U}$ is the set of all time series of finite length.
\item  Constrained form: $\S{U}$ is the set of all time series of length $n$. 
\end{enumerate}
A sample mean is guaranteed to exist in either case but may not be unique \cite{Jain2020}. In addition, computing a sample mean is NP-hard \cite{Bulteau2018}. Consequently, there is an ongoing research on devising heuristics for minimizing the Fr\'echet function. Most contributions focus on devising and applying heuristics for the constrained sample mean problem. State-of-the-art algorithms are stochastic subgradient methods \cite{Schultz2018}, majorize-minimize algorithms \cite{Hautamaki2008,Petitjean2011}, and soft-DTW \cite{Cuturi2017}.  In contrast, only few work has been done for solving the unconstrained sample mean problem. One algorithm is an (essentially optimal) dynamic program that exactly solves the unconstrained problem in exponential time \cite{Brill2019}. A second algorithm is a heuristic, called adaptive DBA (ADBA)~\cite{Liu2019}. This algorithm uses a majorize-minimize algorithm (DBA) as a base-algorithm and iteratively refines subsequences to improve the solution quality.

Currently, there is no clear understanding of the characteristic properties, advantages, and disadvantages of both types of sample means. We can approach the sample mean problem theoretically and empirically. A prerequisite for an empirical approach towards a better understanding of the sample mean problem are sufficiently powerful averaging algorithms. Compared to the constrained sample mean problem, algorithms for the unconstrained sample mean problem are underdeveloped.

In this work, we propose a generic average-compress (AC) algorithm for the unconstrained sample mean problem. The algorithm repeatedly alternates between an averaging (A-step) and a compression (C-step). The A-step requires a time series as initial guess, minimizes the Fr\'echet function, and returns an approximate solution as output. The C-step compresses the approximation of the A-step to obtain an improved solution. The compressed solution of the C-step serves as initial guess of the A-step in the next iteration. Compression is motivated by empirical observations that an unconstrained sample mean is typically shorter than the sample time series to be averaged \cite{Brill2019}. In principle, any averaging algorithm and any compression method can be applied. Here, we propose a compression method that minimizes the squared DTW error between original and compressed time series. Empirical results suggest that the AC scheme substantially outperforms state-of-the-art heuristics including~ADBA. 

This article is organized as follows: Section~\ref{sec:AC} describes the AC algorithm. In Section~\ref{sec:exp} we present and discuss empirical results. Finally, Section~\ref{sec:conc} concludes with a summary of the main findings and an outlook for future research.

\section{Average-Compress Algorithm}\label{sec:AC}

In this section, we develop an average-compress (AC) algorithm for approximately solving the unconstrained sample mean problem. To this end, we first introduce the DTW-distance (Section \ref{subsec:dtw}), the concept of a sample mean under DTW (Section \ref{subsec:sample-mean}), and compressions (Section \ref{subsec:compressions}). Thereafter, we describe the AC algorithm in Section \ref{subsec:ac}. 

\subsection{Dynamic Time Warping} \label{subsec:dtw}

For a given $n \in \N$, we write $[n] = \cbrace{1, \ldots, n}$. A \emph{time series} is a sequence $x = (x_1, \ldots, x_n)$ with elements $x_i \in \R$ for all $i \in [n]$. We denote the length of time series $x$ by $\abs{x} = n$, the set of time series of length $n$ by $\S{T}_n$, and the set of all time series of finite length by $\S{T}$. Consider the ($m \times n$)-grid defined as
\[
[m] \times [n] = \cbrace{(i,j) \,:\, i \in [m], j \in [n]}.
\]
A \emph{warping path} of order $m \times n$ and length $\ell$ is a sequence $p = (p_1 , \dots, p_\ell)$ through the grid $[m] \times [n]$ consisting of $\ell$ points $p_l = (i_l,j_l) \in [m] \times [n]$ such that
\begin{enumerate}
\item $p_1 = (1,1)$ and $p_\ell = (m,n)$
\item $p_{l+1} - p_{l} \in \cbrace{(1,0), (0,1), (1,1)}$ for all $l \in [\ell-1]$. 
\end{enumerate}
The first condition is the boundary condition and the second condition is the step condition of the DTW-distance.  We denote the set of all warping paths of order $m \times n$ by $\S{P}_{m,n}$.
Suppose that $p = (p_1, \ldots, p_\ell) \in \S{P}_{m,n}$ is a warping path with points $p_l = (i_l, j_l)$ for all $l \in [\ell]$. Then $p$ defines an \emph{expansion} (or warping) of the time series $x = (x_1, \ldots, x_m)$ and $y = (y_1, \ldots, y_n)$ to the length-$\ell$ time series $\phi_p(x) = (x_{i_1}, \ldots, x_{i_\ell})$ and $\psi_p(y) = (y_{j_1}, \ldots, y_{j_\ell})$. By definition, the length $\ell$ of a warping path satisfies $\max(m,n) \leq \ell \le m+n$.

The \emph{cost} of warping time series $x$ and $y$ along warping path $p$ is defined by
\begin{equation*}
C_p(x, y) = \normS{\phi_p(x)-\psi_p(y)}{^2} = \sum_{(i,j) \in p} \argsS{x_i-y_j}{^2},
\end{equation*}
where $\norm{\cdot}$ denotes the Euclidean norm and $\phi_p$ and $\psi_p$ are the expansions defined by $p$. The \emph{DTW-distance} of $x$ and $y$ is
\begin{align*}
\dtw(x, y) = \min \cbrace{\sqrt{C_p(x, y)} \,:\, p \in \S{P}_{m,n}}.
\end{align*}
A warping path $p$ with $C_p(x, y) = \dtw^2(x, y)$ is called an \emph{optimal warping path} of $x$ and $y$. By definition, the DTW-distance minimizes the Euclidean distance between all possible expansions that can be derived from warping paths. Computing the DTW-distance and deriving an optimal warping path is usually solved via dynamic programming \cite{Sakoe1978,Vintsyuk1968}. 

\subsection{Sample Means under DTW} \label{subsec:sample-mean}

Let $\S{S} = \cbrace{x_1, \dots, x_N}$ be a sample of $N$ time series $x_i \in \S{T}$. Note that $\S{S}$ is a multiset that allows multiple instances of the same elements.   A sample mean of $\S{S}$ is any time series that minimizes the Fr\'echet function \cite{Frechet1948}
\[
F: \S{U} \rightarrow \R, \quad z \mapsto \frac{1}{N}\sum_{i=1}^N \dtw\!\argsS{z,x_i}{^2},
\]
where $\S{U} \subseteq \S{T}$ is a subset of time series. The value $F(z)$ is the \emph{Fr\'echet variation} of sample $\S{S}$ at $z$. The infimum $\inf_z F(z)$ serves as a measure of variability of $\S{S}$. Here, the search space $\S{U}$ takes one of the following two forms: (i) $\S{U} = \S{T}$ and (ii) $\S{U} = \S{T}_m$. We refer to (i) as the unconstrained and to (ii) as the constrained sample mean problem. Note that the constrained formulation only restricts the length of the candidate solutions, whereas there is no length restriction on the sample time series to be averaged. 

A sample mean exists in either case but is not unique in general \cite{Jain2020}. This result implies that $F$ attains its infimum (has a unique minimum). However, computing a sample mean is NP-hard \cite{Bulteau2018}. The implication is that we often need to resort to heuristics that return useful solutions within acceptable time. 

We briefly describe two state-of-the-art algorithms for the constrained sample mean problem: a stochastic subgradient method (SSG) \cite{Schultz2018} and a majorize-minimize algorithm (DBA) \cite{Hautamaki2008,Petitjean2011}. For a detailed description of both algorithms, we refer to \cite{Schultz2018}. 

To present the update rule of both algorithms in a compact form, we introduce the notions of warping and valence matrix as proposed by \cite{Schultz2018}. Suppose that $p \in \S{P}_{m,n}$ is a warping path. The \emph{warping matrix} of $p$ is the zero-one matrix $W = (w_{ij})\in \{0,1\}^{m \times n}$ with elements
\begin{equation*}
w_{ij} = \begin{cases} 
1 & (i,j) \in p \\ 
0 & \text{otherwise} 
\end{cases}.
\end{equation*}
The \emph{valence matrix} of warping path $p$ is the diagonal matrix $V=(v_{ij}) \in \N^{m \times m}$ with positive diagonal elements
\begin{align*}
v_{ii} = \sum_{j=1}^n w_{ij}.
\end{align*}
Suppose that $z$ and $x$ are time series of length $\abs{z} = m$ and $\abs{x} = n$. Then $W$ warps $x$ onto the time axis of $z$. Each diagonal element $v_{ii}$ of $V$ counts how many elements of $x$ are warped to element $z_i$. 

\begin{algorithm}[t]
\footnotesize 
\caption{Stochastic Subgradient Method}
\begin{algorithmic}[1]
\Procedure{SSG}{$\eta$, $m$, $x_1,\dots,x_N$}
\State initialize solution $z \in \S{T}_m$
\State initialize best solution $z_* = z$
\Repeat
\State{reshuffle order of sample time series}
\For{i}{1}{N} 
\State{compute optimal warping path $p_i$ of $z$ and $x_i$}
\State{compute valence matrix $V_i$ of $p_i$}
\State{compute warping matrix $W_i$ of $p_i$}
\State{update solution $z$ according to the rule\\
\vspace{0.5ex}
\hspace{2cm}$\begin{aligned}
z \gets z - 2 \eta \args{V_i z - W_i x_i}&
\end{aligned}$
}
\EndFor
\State{record best solution $z_* = \argmin\cbrace{F(z_*), F(z)}$}
\Until{termination} 
\State{\Return $z_*$}
\EndProcedure
\end{algorithmic}
\label{alg:ssg}
\end{algorithm}

\paragraph*{Stochastic Subgradient Algorithm.}\ 
Subgradient methods for time series averaging have been proposed by \cite{Schultz2018}. Algorithm \ref{alg:ssg} outlines a vanilla version of the SSG algorithm with constant learning rate $\eta$. In practice, more sophisticated stochastic subgradient variants such as Adam \cite{Kingma2015} are preferred. The input of Algorithm \ref{alg:ssg} are  a learning rate $\eta$, a length-parameter $m$ of the constrained search space, and a sample $x_1,\dots,x_N$ of time series to be averaged. The output is a time series with lowest Fr\'echet variation that has been encountered during optimization. \hfill $\square$

\paragraph*{Majorize-Minimize Algorithm.}\ 
Majorize-minimize algorithms for time series averaging have been proposed in the 1970s mainly by Rabiner and his co-workers with speech recognition as the primary application \cite{Rabiner1979,Wilpon1985}. The early approaches fell largely into oblivion and where successively rediscovered, consolidated, and improved in a first step by Abdulla et al.~\cite{Abdulla2003} in 2003 and then finalized in 2008 by Hautamaki et al.~\cite{Hautamaki2008}.  In 2011, Petitjean et al.~\cite{Petitjean2011} reformulated, explored, and popularized the majorize-minimize algorithm by Hautamaki et al.~\cite{Hautamaki2008} under the name DTW Barycenter Averaging (DBA).

Algorithm \ref{alg:dba} describes the DBA algorithm. It takes a length-parameter $m$ and a sample of time series as input and returns the candidate solution of the last iteration as output. The DBA algorithm terminates after a finite number of iterations in a local minimum of the Fr\'echet function~\cite{Schultz2018}. \hfill $\square$
\begin{algorithm}[t]
\caption{DBA Algorithm}
\begin{algorithmic}[1]
\footnotesize 
\Procedure{DBA}{$m$, $x_1,\dots,x_N$}
\State{initialize solution $z \in \S{T}_m$}
\Repeat
\State{\texttt{//*** Majorize}}
\For{i}{1}{N}
\State{compute optimal warping path $p_i$ of $z$ and $x_i$}
\State{compute valence matrix $V_i$ of $p_i$}
\State{compute warping matrix $W_i$ of $p_i$}
\EndFor
\State{\texttt{//*** Minimize}}
\State{update solution $z$ according to the rule\\
\vspace{0.5ex}
\hspace{2cm}
$\begin{aligned}
z \gets \argsS{\sum_{i=1}^N V_i}{^{\!-1}} \args{\sum_{i=1}^N W_i x_i}
\end{aligned}$
}
\vspace{0.5ex}
\Until{termination} 
\State{\Return $z$}
\EndProcedure
\end{algorithmic}
\label{alg:dba}
\end{algorithm}

\subsection{Compressions}\label{subsec:compressions}

Let $x \in \S{T}$ be a time series of length $n$. A compression of $x$ is a time series $x'$ of length $m \leq n$ that maintains some desirable problem-specific properties of $x$. By definition, $x$ is also a compression of itself. A \emph{compression chain} of $x$ is a sequence $\S{C}(x) = \args{x'_1, \ldots, x'_k}$ of $k \in [n]$ compressions $x'_i$ of $x$ such that 
\[
1 \leq \abs{x'_1} < \abs{x'_2} < \cdots < \abs{x'_k} \leq n.
\]
There are numerous compression methods such as principal component analysis, discrete Fourier transform, discrete wavelet transform, and many more. Here, we consider two simple methods: adaptive scaling (ADA) and  minimum squared DTW error (MSE).

\paragraph*{Adaptive Scaling.}\  
Algorithm \ref{alg:as} describes ADA. The procedure takes a time series $x = (x_1, \ldots, x_n)$ as input and returns  a compression chain $\S{C}(x)$ consisting of $n$ compressions of $x$ of length $1$ to~$n$. To compress a time series $x'_{k+1}$ of length $k+1$ to a time series $x'_k$ of length $k$, ADA merges two consecutive elements  with minimal distance. The merge subroutine in Algorithm \ref{alg:as} replaces these two consecutive time points by their average. 

Finding the smallest distance in Line~\ref{argmin} takes $O(\abs{x'})$ time. In each iteration, the length of $x'$ is reduced by one. Thus, the complexity of computing all $n$ compressions is $O(n^2)$. \hfill $\square$

\begin{algorithm}[t]
\footnotesize
\caption{Adaptive Scaling}
\begin{algorithmic}[1]
\Procedure{ADA}{$x$}
\State $x' \gets x$ \algorithmiccomment{8}{current compression} 
\Repeat
\State $i \in \argmin \{|x'_j - x'_{j+1}| \,:\, j < \abs{x}\}$\label{argmin}
\State $x' \gets$ \textsc{merge}$(x', i)$
\State $\S{C}(x) \gets \S{C}(x) \cup \cbrace{x'}$
\Until{$|x'| = 1$}
\State \textbf{return} $\S{C}(x)$
\EndProcedure
\\
\Procedure{merge}{$x'$, i}
\State $z  \gets x'$
\State replace $z_i$ by $(x'_i+x'_{i+1})/2$
\State delete $z_{i+1}$
\State \textbf{return} $z$
\EndProcedure
\end{algorithmic}
\label{alg:as}
\end{algorithm}

\paragraph*{Minimum Squared DTW Error Compression.}\  
The second compression method computes a time series of a given length such that the squared DTW error is minimized. Let $x \in \S{T}$ be a time series of length $n$ and let $m< n$. We call each
\[
x' \in \argmin \cbrace{\dtw\!\argsS{x, z}{^2} \,:\, z \in \S{T}_m}
\]
an MSE compression of $x$ of length $m$. Observe that the MSE compression problem for $x$ is the constrained sample mean problem of the sample $\S{S} = \cbrace{x}$.
Algorithm~\ref{alg:mse} outlines MSE compression.

It is not hard to see that for a compression~$x'$, an optimal warping path~$p$ between~$x$ and~$x'$ warps every element of~$x$ to exactly one element of~$x'$, that is, $\phi_p(x)=x$.
Thus, we can write
\begin{align} \label{eq:form_of_p}
\begin{split}
p&=\big((1,1),\ldots,(i_1,1),(i_1+1,2),\ldots,\\
&\phantom{=((}(i_1+i_2,2),\ldots,(n-i_m),\ldots,(n,m)\big),  
\end{split}
\end{align}
where $\sum_{l=1}^mi_l = n$. Let~$d_l=\sum_{j=1}^li_j$, $l\in[m]$, and~$d_0=0$. The squared DTW error of~$x'$ is
\begin{align}\label{eq:mse-cost}
\dtw(x,x')^2 =\sum_{l=1}^m\sum_{i=d_{l-1}+1}^{d_l} (x_i-x'_l)^2.
\end{align}

MSE compression is also known as \emph{adaptive piecewise constant approximation}~\cite{Chakrabarti2002} and as \emph{segmentation} problem \cite{Terzi2006}.
It can be solved exactly via dynamic programming in $O(n^2m)$ time \cite{Bellman1962}. Moreover, the dynamic program allows to find all $n$ compressions (for each length $m =1,\dots,n$) in $O(n^3)$ time by running it once for $m=n$ (as it is done in Algorithm~\ref{alg:mse}).

Interestingly, MSE compression is also related to one-dimensional $k$-means clustering. To see this relationship, consider an optimal warping path between the compression~$x'$ and~the original time series $x$ as in Eq.~\eqref{eq:form_of_p}. Then, the squared DTW error is minimal for~$x'_l= (x_{d_{l-1}+1} + \cdots + x_{d_l})/i_l$. Thus, finding an MSE compression $x'$ of length $k$ can also be seen as a one-dimensional $k$-means clustering problem, where every cluster consists of a \emph{consecutive} subsequence of elements in~$x$. Indeed, the dynamic program in Algorithm~\ref{alg:mse} is the same as for one-dimensional $k$-means~\cite{Wang2011} (without previously sorting the elements in~$x$). 

To reduce the computational complexity, several heuristics and approximations for MSE compression have been proposed \cite{Terzi2006,Keogh2001,Yi2000,Chakrabarti2002}. Also ADA compression can be regarded as a heuristic for MSE compression since it greedily averages two consecutive elements. 
\hfill $\square$

\bigskip

To conclude, with MSE compression, we consider an exact solution method to a sound compression problem and with ADA compression, we consider a fast heuristic.
Among the various heuristics, we have chosen ADA compression because it has been successfully tested for improving approximate solutions of the constrained sample mean problem~\cite{Petitjean2011}.

\begin{algorithm}[b]
\footnotesize
\caption{MSE Compression}
\begin{algorithmic}[1]
  \Procedure{MSE}{$x$}
  \State initialize tables~$D$, $C$
  \State $C[1,1]\gets (x_1)$
  \State $D[1,1] \gets 0$
  \For{i}{2}{|x|}
    \For{m}{1}{i}
      
      \For{j}{m}{i}
        \State $\mu \gets (\sum_{l=j}^ix_l) / (i-j+1)$
        \State $d \gets \sum_{l=j}^i(x_l-\mu)^2$
        \If {$D[j-1,m-1]+d < D[i,m]$}
          \State $D[i,m] \gets D[j-1,m-1]+d$
          \State $C[i,m] \gets C[j-1,m-1].\text{append}(\mu)$
        \EndIf
      \EndFor
    \EndFor
  \EndFor
  \State $\S{C}(x) \gets \{C[|x|,m] \,:\, 1\le m\le|x|\}$
  \State \textbf{return} $\S{C}(x)$
  \EndProcedure
\end{algorithmic}
\label{alg:mse}
\end{algorithm}

\subsection{The Average-Compress Algorithm} \label{subsec:ac}

\begin{algorithm}[t]
\footnotesize
\caption{Average-Compress Algorithm}
\begin{algorithmic}[1]
\Procedure{AC}{$\S{X}, z$}
\State $z_* \leftarrow z$
	\algorithmiccomment{4.15}{best solution found so far} 
\State $f_* \leftarrow F(z_*)$
	\algorithmiccomment{2}{variation of $z_*$} 
\State $\ell_* \leftarrow \abs{z_*}$
	\algorithmiccomment{3.15}{length of $z_*$} 
\Repeat
\State A-Step: $z \leftarrow \textsc{average}(\S{X}, z)$
\label{alg:mean-compression:mean}
\State C-Step: $\S{C}(z) \leftarrow \textsc{compress}(z)$
\label{alg:mean-compression:compress}
\State{\texttt{//*** Evaluate solution}}
\State $z \leftarrow \argmin\cbrace{F(z') \,:\, z' \in \S{C}(z) \cup \cbrace{z}}$
\If{$F(z) < f_*$ \textbf{or} ($F(z) = f_*$ \textbf{and} $\abs{z} < \ell_*$)}
\State $f_* \leftarrow F(z)$
\State $l_* \leftarrow \abs{z}$
\State $z_* \leftarrow z$
\EndIf
\Until{convergence}
\State \textbf{return} $z_*$
\EndProcedure
\end{algorithmic}
\label{alg:mean-compression}
\end{algorithm}

In this section, we assemble the pieces of the previous sections and propose a generic average-compress (AC) algorithm for approximately solving the unconstrained sample mean problem. 

\paragraph*{AC Algorithm.}\  
The AC algorithm alternates between averaging (A-step) and compression (C-step). For this purpose, any averaging algorithm and any compression method can be used. Algorithm~\ref{alg:mean-compression} depicts the generic procedure. The input of the algorithm is a sample $\S{S}$ of time series and an initial guess $z \in \S{T}$. It then repeatedly applies the following steps until termination:
\begin{enumerate}
\itemsep0em
\item A-step: approximate sample mean 
\[
z \gets  \textsc{average} (\S{S}, z).
\]
\item C-step: compute a compression chain  
\[
\S{C}(z)\gets \textsc{compress}(z).
\]
\item Evaluate solution: 
\begin{enumerate}
\item Select the shortest compression $z_* \in \S{C}(z)$ such that $F(z_*) \leq F(z')$ for all $z' \in \S{C}(z)$. 
\item If $F(z_*) < F(z)$, set $z \gets z_*$ and go to Step~1, otherwise terminate.
\end{enumerate}
\end{enumerate}
Line~\ref{alg:mean-compression:compress} computes the complete compression chain $\S{C}(z)$ that consists of all $\abs{z}$ compressions of $z$ of lengths $1$ to $\abs{z}$. To accelerate the algorithm at a possible expense of solution quality, sparse compression chains can be considered. \hfill $\square$

\medskip

In the following, we explain why and under which conditions compression is useful. To simplify our argument, we assume that AC uses an averaging algorithm for the constrained sample mean problem (such as SSG or DBA).  In this case, the length $m$ of the initial guess restricts the search space of AC to the set $\S{T}_{\leq m}$ of all time series of maximum length $m$. The choice of the length-parameter $m$ via the initial guess is critical. If $m$ is too small, the search space $\S{T}_{\leq m}$ may not contain an unconstrained sample mean.
For a given sample $\S{S}$, the Reduction Theorem~\cite{Jain2020} guarantees the existence of an unconstrained sample mean of a length at most $m_{\S{S}} = \sum_{x\in\S{S}}|x| - 2(|\S{S}|-1)$. Consequently, we can safely constrain the search space to $\S{T}_{\leq m_{\S{S}}}$ for solving the unconstrained sample mean problem.
Then, a naive approach to minimize the Fr\'echet function on $\S{T}_{\leq m_{\S{S}}}$ is to solve $m_{\S{S}}$ constrained sample mean problems on $\S{T}_{m_{\S{S}}}, \ldots, \S{T}_1$ and then to pick the solution with lowest Fr\'echet variation. When using state-of-the-art heuristics for the $m_{\S{S}}$ constrained problems, the naive approach is computationally infeasible.

The purpose of compression is to substantially accelerate the intractable naive approach at the expense of solution quality. Instead of solving all $m_{\S{S}}$ constrained problems, the AC algorithm uses compressions to select a few promising search spaces $\S{T}_{m_0}, \S{T}_{m_1}, \ldots, \S{T}_{m_k}$ with $m_{\S{S}} = m_0 > m_1 > \cdots > m_k \geq 1$.  Starting with $\S{T}_{m_{\S{S}}} = \S{T}_{m_0}$, the solution $z_{i-1}$ found in $\S{T}_{m_{i-1}}$ is compressed in order to determine the next search space $\S{T}_{m_i}$. The length-parameter $m_i$ of the next search space $\S{T}_{m_i}$ corresponds to the length of the compression $z_i$ of $z_{i-1}$ with lowest Fr\'echet variation. Obviously, this idea only accelerates the naive approach if the length~$m_i$ of the best compression is substantially smaller than the length $m_{i-1}$ of the previous solution. 

The theoretical upper bound~$m_{\S{S}}$ provided by the Reduction Theorem~\cite{Jain2020} is usually very large such that existing state-of-the-art heuristics for solving the constrained problem on $\S{T}_{\leq m_{\S{S}}}$ are computationally intractable. In this case, also the AC algorithm using such a heuristic would be infeasible. However, empirical results on samples of two time series of equal length~$n$ suggest that the length of an unconstrained sample mean is more likely to be less than~$n$ \cite{Brill2019}. Similar results for larger sample sizes are unavailable due to forbidding running times required for exact sample means. For solving constrained sample mean problems, it is common practice to choose $m$ within the range of the lengths of the sample time series \cite{Petitjean2011}.  Within this range, experimental results showed that an approximate solution of a constrained sample mean can be improved by reducing its length using adaptive scaling~\cite{Petitjean2011}. These findings suggest to choose the length-parameter $m$ within or slightly above the range of lengths of the sample time series.

\section{Experiments}\label{sec:exp}
\newcommand{\dba}{\text{DBA}}
\newcommand{\dbaada}{DBA-ADA}
\newcommand{\dbamse}{DBA-MSE}
\newcommand{\ssg}{SSG}
\newcommand{\ssgmse}{SSG-MSE}
\newcommand{\adba}{ADBA}
\newcommand{\adbamse}{ADBA-MSE}
\newcommand{\edp}{\text{EDP}}

Our goal is to assess the performance of the proposed AC algorithm. For our experiments, we use the~$85$ data sets from the UCR archive \cite{Chen2015}. Appendix~\ref{app:subsec:hyper-parameters} summarizes the parameter settings of the mean algorithms used in these experiments.

\subsection{Comparison of ADA and MSE}

We compared the performance of ADA and MSE as compression subroutines of the AC algorithm.
We applied the following configurations:  
\begin{center}
\begin{footnotesize}
\begin{tabular}{llc}
\toprule
Acronym & Algorithm & $I$\\
\midrule
\dba & DTW Barycenter Averaging \cite{Hautamaki2008,Petitjean2011} & --\\
\dbaada$_1$ & DBA with ADA compression  \cite{Petitjean2011} & 1\\
\dbamse$_1$ & DBA  with MSE compression & 1\\
\dbaada &  DBA with ADA compression & $*$\\
\dbamse & DBA with  MSE compression & $*$\\
\bottomrule
\end{tabular}
\end{footnotesize}
\end{center}
Column $I$ refers to the number of iterations of the repeat-until loop of the AC Algorithm. Compression schemes with $*$ iterations run until convergence. We applied DBA and  the four AC algorithms to approximate the class means of every UCR training set.\footnote{The UCR data sets have prespecified training and test sets.}
 
To assess the performance of the mean algorithms, we recorded the percentage deviations, ranking distribution, and space-saving ratios. Here, we used the solutions of the DBA algorithm as reference. The percentage deviation of a mean algorithm~$A$ is defined by
\[
p_{\text{dev}}(A) = 100\cdot\frac{F(z_A)-F(z_{\dba})}{F(z_{\dba})},
\]
where $z_{\dba}$ is the solution of the DBA algorithm and $z_A$ is the solution of algorithm~$A$. Negative (positive) percentage deviations mean that algorithm~$A$ has better (worse) Fr\'echet variation than DBA. The ranking distribution summarizes the rankings of every mean algorithm over all samples. The best (worst) algorithm is ranked first (last). The space-saving ratio of algorithm~$A$ is 
\[
\rho_{\text{ss}}(A) = 1 - \frac{\abs{z_A}}{\abs{z_{\dba}}}.
\] 
A positive (negative) space-saving ratio means that the solution $z_A$ is shorter (longer) than $z_{\dba}$.

Table~\ref{tab:res_bp_mean_01} summarizes the results. The top table shows the average, standard deviation, minimum, and maximum percentage deviations from the Fr\'echet variation of the \dba~algorithm (lower is better). The table in the middle shows the distribution of rankings and their corresponding averages and standard deviations. The best (worst) algorithm is ranked first (fifth). Finally, the bottom table shows the average, standard deviation, minimum, and maximum space-saving ratios (higher is better).

All AC variants improved the solutions of the \dba~baseline by $4.6 \%$ to $7.0 \%$ on average and $45\% (\pm 2\%)$  in the best case. By construction, an AC solution is never worse than a \dba~solution. The best method is \dbamse~with average rank $1.0$ followed by \dbamse$_1$ and \dbaada~with average ranks $2.4$ and $2.5$, respectively. These three methods clearly outperformed \dbaada$_1$ proposed by Petitjean et al.~\cite{Petitjean2011}. The main improvement of \dbamse~and \dbaada~occurs at the first iteration.

We considered the lengths of the approximated means. Recall that all mean algorithms started with the same initial guess (medoid). The length of a \dba~solution corresponds to the length of its initial guess, whereas solutions of AC algorithms are likely to be shorter by construction. The bottom table shows that solutions of ADA compression save about a third ($0.34$, $0.36$) of the length of \dba~solutions on average and solutions of MSE compressions close to a half ($0.44$, $0.45$). As for the Fr\'echet variation, most of the space-saving occurs in the first iteration of an AC algorithm.

\begin{table*}[t]
\caption{Results of ADA and MSE compressions.}
\label{tab:res_bp_mean_01}
\centering
\small
\begin{tabular}[t]{lrrrrr}
\toprule
\multicolumn{6}{c}{\textbf{Percentage deviations}}\\
\midrule
\phantom{rank}&  \dba & \dbaada$_1$ & \dbamse$_1$ &  \dbaada & \dbamse\\
\midrule
avg & 0.0 & -4.6 & -6.0  & -5.6  & -7.0\\
std & 0.0 & 4.8 & 5.9 & 5.6 & 6.6\\
min & 0.0 & -43.1 & -43.9 &  -47.1 & -47.7\\
max & 0.0 & 0.0 & 0.0 & 0.0  & 0.0\\
\bottomrule
\\
\toprule
\multicolumn{6}{c}{\textbf{Ranking distribution}}\\
\midrule
Rank & \dba & \dbaada$_1$ & \dbamse$_1$ &  \dbaada & \dbamse\\
\midrule
1&0.5&2.5&2.7&10.2&96.8\\
2&0.2&0.6&56.2&34.6&3.2\\
3&0.3&9.7&37.0&55.2&0.0\\
4&0.0&87.1&4.1&0.0&0.0\\
5&99.0&0.0&0.0&0.0&0.0\\
\midrule
avg&5.0&3.8&2.4&2.5&1.0\\
std&0.3&0.6&0.6&0.7&0.2\\
\bottomrule
\\
\toprule
\multicolumn{6}{c}{\textbf{Space-saving ratios}}\\
\midrule
 & \dba & \dbaada$_1$ & \dbamse$_1$ &  \dbaada & \dbamse\\
\midrule
avg & 0.00 & 0.34 & 0.44 & 0.36 & 0.45\\
std & 0.00 & 0.21 & 0.24 & 0.22 & 0.24\\
min & 0.00 & 0.00 & 0.00 & 0.00 & 0.00\\
max & 0.00 & 0.96 & 0.99 & 0.96 & 0.99\\
\bottomrule
\end{tabular}
\end{table*}

\subsection{Comparison of AC Algorithms}

The goal of the second experiment is to compare the performance of the following mean algorithms:
\begin{center}
\begin{footnotesize}
\begin{tabular}{llc}
\toprule
Acronym & Algorithm & $I$\\
\midrule
\dba & DTW Barycenter Averaging \cite{Hautamaki2008,Petitjean2011} & --\\
\ssg & stochastic subgradient method \cite{Schultz2018}& --\\
\adba & adaptive DBA \cite{Liu2019} & --\\
\dbamse& DBA with MSE compression & $*$\\
\ssgmse& SSG with MSE compression& $*$\\
\adbamse& adaptive DBA with MSE compression & $*$\\
\bottomrule
\end{tabular}
\end{footnotesize}
\end{center}

Table~\ref{tab:res_bp_mean_02} summarizes the results using the same legend as in Table \ref{tab:res_bp_mean_01}. The percentage deviations and rankings suggest that the three AC variants \dbamse, \ssgmse, and \adbamse~performed substantially better than the corresponding base algorithms \dba, \ssg, and \adba, respectively. The \ssgmse~algorithm performed best with an average rank of $1.4$, followed by \adbamse~($2.9$), \ssg~($3.1$), and \dbamse~($3.3$). 
Interestingly, \adba~performed worse than \dbamse.
Both, \adba~and \dbamse, are based on the \dba~algorithm. The difference between both algorithms is that \adba~compresses and expands selected subsequences of the current \dba~solution, whereas \dbamse~only compresses the current \dba~solution. The results indicate that simple MSE compression on the entire sequence appears to be a better strategy than \adba's compression and expansion schemes on selected subsequences. 
Notably, \ssg~performed best among the three base averaging algorithms \dba, \ssg, \adba, and performed even better than \dbamse. These results are in contrast to those presented in \cite{Liu2019}, where \adba~outperformed \ssg~(and also \dba). Our findings confirm that the performance of \ssg~substantially depends on a careful selection of an optimizer (such as Adam) and a proper choice of the initial learning rate. 

Next, we examine the length of the solutions. Note that \ssg~also does not alter the length of its initial guess such that $\rho_{\text{ss}}(\text{DBA}) = \rho_{\text{ss}}(\text{SSG}) = 0$. The bottom table shows that MSE compression schemes reduce the length of the solutions obtained by their corresponding base algorithm (\dba, \ssg, and \adba). The space-saving ratios of the AC variants are roughly independent of the particular base algorithm for mean computation (0.43--0.45). Notably, the base algorithm \adba~is more likely to compress rather than to expand the \dba~solutions. This finding is in line with the hypothesis that an exact mean is typically shorter than the length of the sample time series \cite{Brill2019}.

\begin{table*}[t]
\caption{Results of the AC algorithms.}
\label{tab:res_bp_mean_02} 
\centering
\small
\begin{tabular}{lrrrrrr}
\toprule
\multicolumn{7}{c}{\textbf{Percentage deviations}}\\
\midrule
& \dba & \ssg & \adba & \dbamse & \ssgmse & \adbamse \\
\midrule
avg & 0.0 &  -7.4 & -6.4 & -7.0 & -11.6 & -9.4\\
std & 0.0 & 16.1 & 10.4 & 6.6 & 13.8 & 10.5\\
min & 0.0 & -79.7 & -81.8 & -47.7 & -83.3 & -82.0\\
max & 0.0 & 188.1  & 17.7  & 0.0 & 50.0 & 13.3\\
\bottomrule
\\
\toprule
\multicolumn{7}{c}{\textbf{Ranking distribution}}\\
\midrule
Rank & \dba & \ssg & \adba & \dbamse & \ssgmse & \adbamse \\
\midrule
1 & 1.9 & 5.1 & 1.1 & 9.5 & 80.6 & 8.9\\
2 & 2.1 & 37.6 & 1.7 & 22.9 & 10.5 & 21.9\\
3 & 2.4 & 20.8 & 8.6 & 25.9 & 2.1 & 41.4\\
4 & 8.1 & 19.5 & 33.7 & 14.0 & 0.5 & 21.0\\
5 & 11.4 & 11.1 & 36.7 & 27.8 & 6.3 & 6.8\\
6 & 74.1 & 5.9 & 18.3 & 0.0 & 0.0 & 0.0\\
\midrule
avg & 5.5 & 3.1 & 4.6 & 3.3 & 1.4 & 2.9\\
std & 1.1 & 1.3 & 1.0 & 1.3 & 1.0 & 1.0\\
\bottomrule
\\
\toprule
\multicolumn{7}{c}{\textbf{Space-saving ratios}}\\
\midrule
& \dba & \ssg & \adba & \dbamse & \ssgmse & \adbamse \\
\midrule
avg & 0.00 &  0.00 & 0.31 & 0.45 & 0.43 & 0.45\\
std & 0.00 &  0.00 & 0.21 & 0.24 & 0.24 & 0.24\\
min & 0.00 &  0.00 & -0.04 & 0.00 & 0.00 & 0.00\\
max & 0.00 &  0.00 & 0.91 & 0.99 & 0.99 & 0.97\\
\bottomrule
\end{tabular}
\end{table*}

\subsection{Qualitative Analysis of Mean Algorithms}

\begin{figure}[t]
\begin{subfigure}{.48\textwidth}
  \centering
  \includegraphics[width=\linewidth]{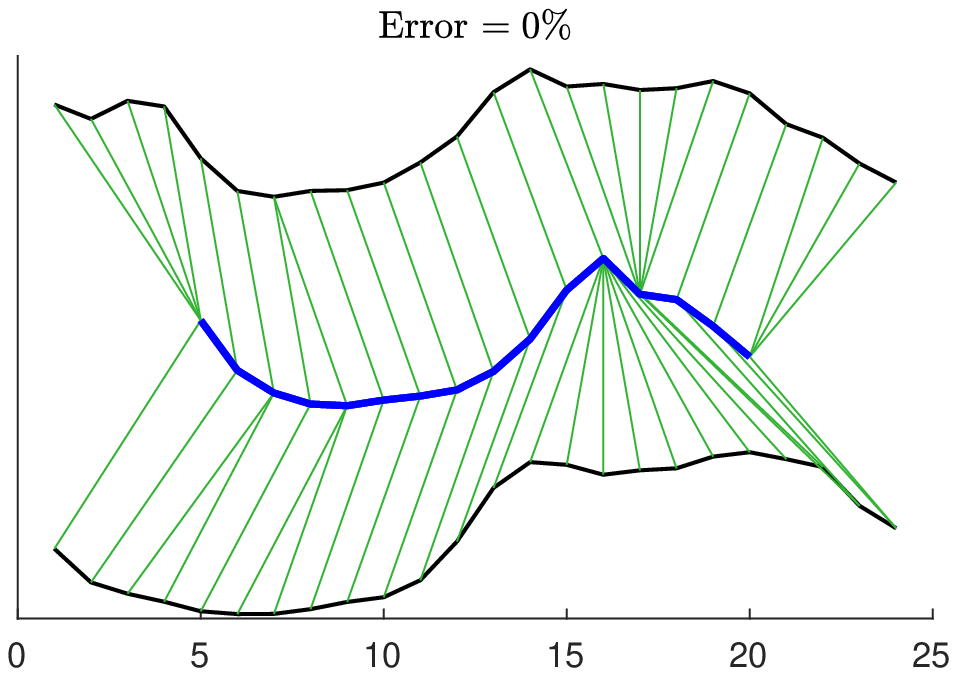}
  \caption{\edp (exact solution)}
  \label{fig:sub1}
\end{subfigure}%
\hfill
\begin{subfigure}{.48\textwidth}
  \centering
  \includegraphics[width=\textwidth]{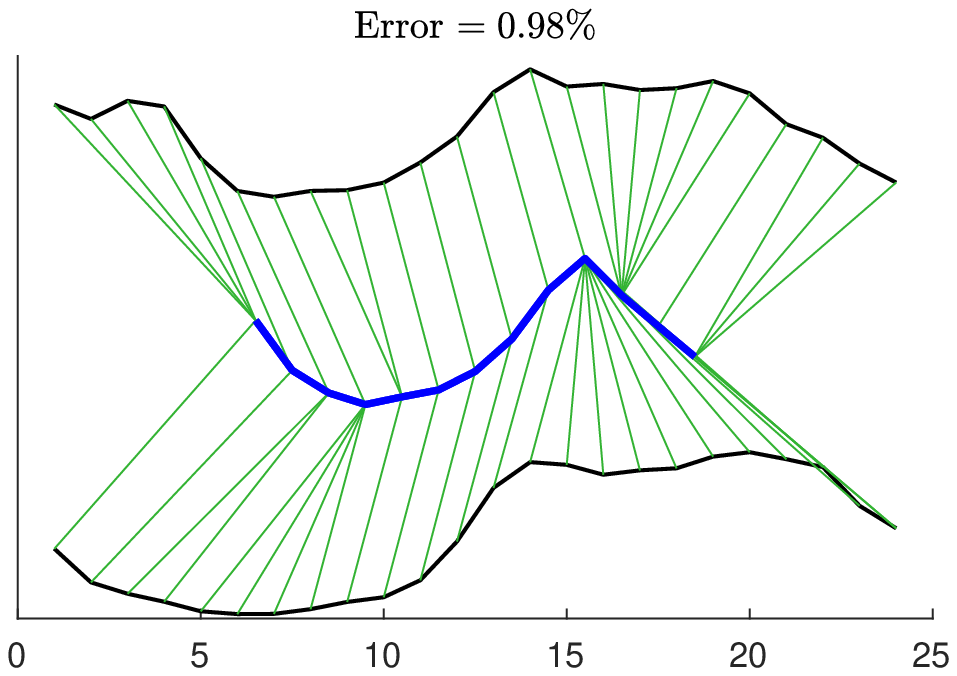}
  \caption{\ssgmse}
  \label{fig:sub2}
\end{subfigure}

\bigskip

\begin{subfigure}{.48\textwidth}
  \centering
  \includegraphics[width=\linewidth]{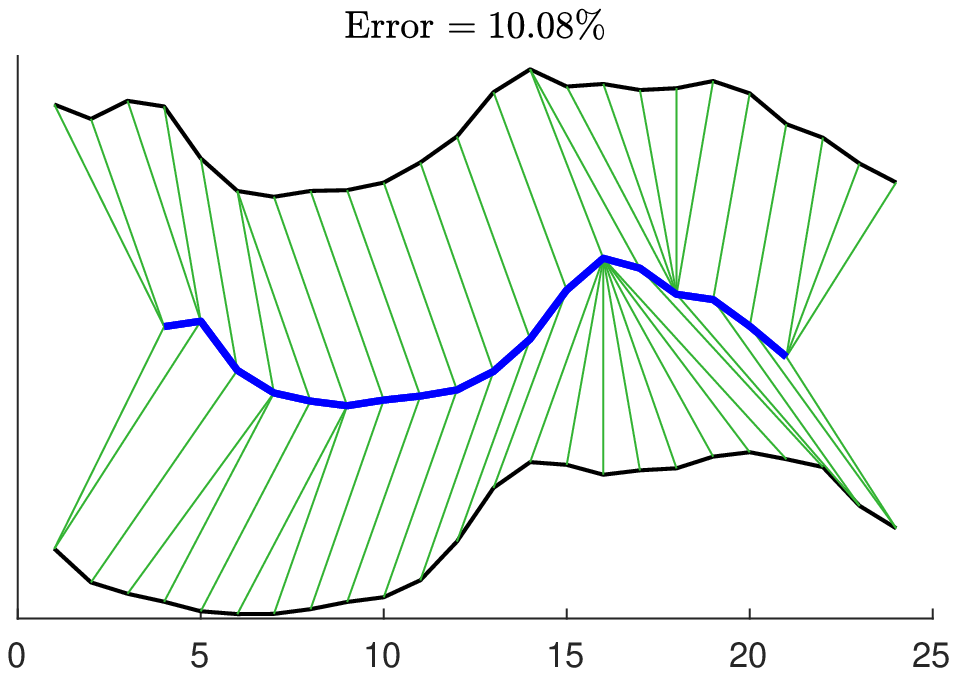}
  \caption{\adba}
  \label{fig:sub3}
\end{subfigure}%
\hfill
\begin{subfigure}{.48\textwidth}
  \centering
  \includegraphics[width=\textwidth]{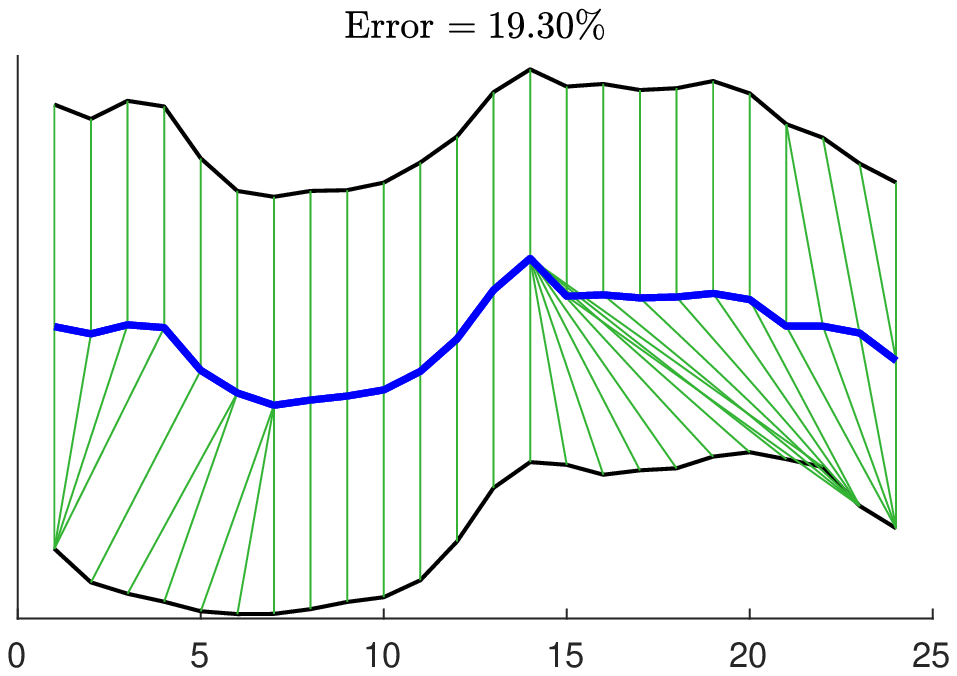}
  \caption{\dba}
  \label{fig:sub4}
\end{subfigure}

\caption{Comparison of different sample mean algorithms. Each plot shows the mean (blue) of two time series of length $24$ (black) and the optimal alignments (green) as found by the corresponding algorithm. The error specified above each plot is the percentage deviation of the Fr\'echet variation of the corresponding solution from the minimal Fr\'echet variation. The length of the means are $16$ in (a), $13$ in (b), $18$ in (c), and $24$ in (d). }
\label{fig:test}
\end{figure}

The goal of this section is to qualitatively analyze the behavior and phenomena behind the different types of mean algorithms. For this, we considered \dba, \adba, and the AC variant \ssgmse ~relative to an exact dynamic program (\edp) proposed by \cite{Brill2019}. Since the sample mean problem is NP-hard, we only considered a sample with two sample time series of length $24$ from the Chinatown data set \cite{Dau2019}. The two sample time series slightly differ in their amplitudes and on a high abstraction level, they have the following features in common: Both start with a wide valley followed by a peak, a flat plateau-like valley at a high altitude until they finally end with a descent. 

Figure \ref{fig:test} shows the sample time series and the sample means returned by the four algorithms. The four algorithms differ in the level of feature abstraction and in susceptibility of spurious features, whereby lower level feature representations are more prone to spurious features than higher level ones: 
The \edp ~exhibits the common shape of both sample time series. In addition, it filters out variations of speed by condensing the mean to length $16$. The \ssgmse ~algorithm more aggressively condenses a solution than \edp ~resulting in a more compact and higher level description of a mean of length $13$. In contrast to \edp, the AC variant \ssgmse ~has smoothed out the flat plateau-like valley. The solution of \adba ~more moderately condenses a solution than \ssgmse ~and \edp ~resulting in a lower level representation of length $18$. In addition, \adba ~includes a spurious plateau at the beginning that occurs only in the upper sample time series. Finally, \dba ~aims at capturing the common features of both sample time series with respect to a predefined length (here $24$). The resulting solution contains more low level features than the other approaches and includes spurious features which occur in only one of both time series. Finally, we hypothesize that spurious features may also occur in exact solutions when two sample time series do not share many common features.

Figure \ref{fig:test} shows the error of each mean algorithm as the percentage deviation of their Fr\'echet variations from the minimum Fr\'echet variation. The Fr\'echet variation measures the amount of dispersion of a sample of time series. Such a measure is, for example, important in evaluating $k$-means clustering using validation indices based on the Fr\'echet variation for each cluster. The errors of the heuristics differ substantially with \edp ~ranked first followed by ~\ssgmse ~($1.0\%$), \adba ~($10.1\%$), and \dba ~($19.3\%$). We hypothesize that low level and spurious features could result in erroneous measures of dispersion. These errors then propagate to pattern recognition methods based on time series averaging such as $k$-means clustering.

\subsection{Application: $k$-Means Clustering}

In this experiment, we investigated how the quality of a mean algorithm affects the quality of a $k$-means clustering.
Let $\S{S} = \cbrace{x_1, \dots, x_n}\subseteq \S{T}$ be a set of $n$ finite time series.
The goal of $k$-means is to find a set $\S{Z} = \cbrace{z_1, \ldots, z_k}$ of $k$ centroids $z_j \in \S{T}$ such that the $k$-means error 
\[
J(\S{Z}) = \frac{1}{n}\sum_{i=1}^n \min_{z \in \S{Z}} \dtw(x_i, z)^2
\]
is minimized. We used \dba, \ssg, \adba, \dbamse, \ssgmse, and \adbamse~for computing the set $\S{Z}$ of centroids. We applied the six variants of $k$-means to $70$ UCR data sets and excluded $15$ UCR data sets due to overly long running times (see Appendix~\ref{app:subsec:k-means}). We merged the prespecified training and test sets. The number $k$ of clusters was set to the number of classes and the centroids were initialized by the class medoids.

Table~\ref{tab:res_kmeans} summarizes the results. The top table presents the average, standard deviation, minimum, and maximum percentage deviations from the respective minimum $k$-means error (lower is better). The percentage deviation of $k$-means algorithm $A$ for data set $D$ is defined by
\[
p_{\text{dev}}(A, D) = 100*\frac{J(\S{Z}_A)- J(\S{Z}_D)}{J(\S{Z}_D)},
\]
where $\S{Z}_A$ is the set of centroids returned by algorithm~$A$ and $\S{Z}_D$ is the best solution obtained by one of the six $k$-means algorithms.  The bottom table shows the distribution of rankings and their corresponding averages and standard deviations. The best (worst) algorithm is ranked first (sixth)

The results show that the AC approach substantially improved all $k$-means variants using one of the base averaging methods (\dba, \ssg, \adba). Notably, \ssgmse\ performed best with an average percentage deviation of $1.5 \%$ and an average rank of $1.4$, followed by \adbamse~($3.8 \%$ and $2.9$). The average percentage deviations of \dba\ and \dbamse\ are substantially impacted by the results on a single data set (DiatomSizeReduction). Removing the DiatomSizeReduction data set yields an average percentage deviation of~$13.2$ for \dba\ and $4.6$ for \dbamse, whereas the other average percentage deviations remain unchanged up to $\pm 0.1 \%$. These findings confirm the hypothesis raised by Brill et al.~\cite{Brill2019} that better mean algorithms more likely result in lower $k$-means errors.  

\begin{table*}[t]
\caption{Results of $k$-means clustering.}
\label{tab:res_kmeans} 
\centering
\small
\begin{tabular}{lrrrrrr}
\toprule
\multicolumn{7}{c}{\textbf{Percentage deviations}}\\
\midrule
& \dba & \ssg & \adba & \dbamse & \ssgmse & \adbamse\\
\midrule
avg & 20.0 & 8.3 & 8.1 & 11.5 & 1.5 & 3.8\\
std & 58.1 & 10.8 & 4.8 & 58.1 & 6.2 & 3.5\\
max & 491.8 & 55.6 & 26.5 & 488.5 & 44.4 & 17.1\\
\bottomrule
\\
\toprule
\multicolumn{7}{c}{\textbf{Ranking distribution}}\\
\midrule
Rank & \dba & \ssg & \adba & \dbamse & \ssgmse & \adbamse\\
\midrule
1 & 2.9 & 2.9 & 0.0 & 14.3 & 81.4 & 5.7\\
2 & 0.0 & 22.9 & 0.0 & 30.0 & 12.9 & 28.6\\
3 & 4.3 & 15.7 & 14.3 & 24.3 & 0.0 & 42.9\\
4 & 7.1 & 21.4 & 41.4 & 11.4 & 0.0 & 18.6\\
5 & 11.4 & 31.4 & 27.1 & 18.6 & 5.7 & 2.9\\
6 & 74.3 & 5.7 & 17.1 & 1.4 & 0.0 & 1.4\\
\midrule
avg & 5.5 & 3.7 & 4.5 & 2.9 & 1.4 & 2.9\\
std & 1.1 & 1.4 & 0.9 & 1.4 & 1.0 & 1.0\\
\bottomrule
\end{tabular}
\end{table*}

\section{Conclusion}\label{sec:conc}

We formulated a generic average-compress algorithm for the unconstrained sample mean problem in DTW spaces. Starting with an initial guess of sufficient length, the AC algorithm alternates between averaging and compression. In principle, any averaging and any compression algorithm can be plugged into the AC scheme. The compression guides the algorithm to promising search spaces of shorter time series. This approach is theoretically justified by the Reduction Theorem \cite{Jain2020} that guarantees the existence of an unconstrained sample mean in a search space of bounded length. Experimental results show that the AC algorithm substantially outperforms state-of-the-art heuristics for time series averaging. In addition, we observed that better averaging algorithms yield lower $k$-means errors on average. Open research questions comprise application of the AC scheme to the empirical analysis of alternative compression methods for the AC algorithm and reducing its computational effort.

\paragraph*{\textbf{Acknowledgement.}}\ 
B.~Jain was funded by the DFG project JA~2109/4-2.

\appendix
\section{Experimental Settings}\label{sec:experimental-details}

\subsection{Hyperparameter Settings}\label{app:subsec:hyper-parameters}
In all experiments, we selected the sample medoid as initial guess of a mean algorithm. The DBA algorithm terminated after convergence and latest after $50$ epochs (cycles through a sample). The ADBA algorithm terminates subsequence optimization  when the sum of the scaling coefficients changes its sign and latest after $50$ iterations. The \ssg~algorithm terminated after $50$ iterations without observing an improvement and latest after $\max\args{50, 5000/n}$ epochs. As optimization scheme, \ssg~applied Adam~\cite{Kingma2015} with $\beta_1 = 0.9$ as first and $\beta_2 = 0.999$ as second momentum. To cope with the problem of selecting an initial learning rate, we used the procedure described in Algorithm \ref{alg:ssg-lr}. The input is a sample $\S{S}$ of size $n$. The output is the best solution found. The algorithm terminates if the solution did not improve for two consecutive learning rates and latest if $\sqrt{n}/2^i \leq 10^{-6}$.
\begin{algorithm}
\footnotesize
\caption{SSG with learning rate selection}
\begin{algorithmic}[1]
\Procedure{SSG}{$\S{S}$}
\State $n \gets \abs{\S{S}}$
\State $i \gets 1$
\Repeat
\State test SSG with learning rate $\sqrt{n}/2^i$
\State record best solution $z_*$ found so far
\State $i \gets i+1$
\Until{convergence}
\State \textbf{return} $z_*$
\EndProcedure
\end{algorithmic}
\label{alg:ssg-lr}
\end{algorithm}

\subsection{Data Sets Excluded From $k$-Means Experiments}\label{app:subsec:k-means}
The following list contains all UCR data sets excluded from $k$-means clustering due to computational reasons:
\begin{center}
\footnotesize
\begin{tabular}{l@{\qquad}l@{\qquad}l}
\toprule
CinCECGtorso & Phoneme \\
FordA & StarLightCurves \\
FordB & UWaveGestureLibraryAll \\
HandOutlines & UWaveGestureLibraryX \\
InlineSkate & UWaveGestureLibraryY \\
Mallat & UWaveGestureLibraryY\\
NonInvasiveFatalECGThorax1 & Yoga\\
NonInvasiveFatalECGThorax2 &\\
\bottomrule
\end{tabular}
\end{center}


\begin{thebibliography}{}
\setlength{\parskip}{0pt}
\setlength{\itemsep}{0pt plus 0.3ex}
\small

\bibitem{Abanda2018}
A.~Abanda, U.~Mori, and J.A.~Lozano. 
\newblock A review on distance based time series classification. 
\newblock \emph{Data Mining and Knowledge Discovery}, 33(2):378--412, 2019.

\bibitem{Abdulla2003}
W.H.~Abdulla, D.~Chow, and G.~Sin. 
\newblock Cross-words reference template for DTW-based speech recognition systems. 
\newblock \emph{Conference on Convergent Technologies for Asia-Pacific Region}, 2003.

\bibitem[Aghabozorgi et al., 2015]{Aghabozorgi2015}
S.~Aghabozorgi, A.S.~Shirkhorshidi, and T.-Y.~Wah.
\newblock Time-series clustering -- A decade review.
\newblock \emph{Information Systems}, 53:16--38, 2015.

\bibitem[Bagnall et al., 2017]{Bagnall2017}
A.~Bagnall, J.~Lines, A.~Bostrom, J.~Large, and E.~Keogh. 
\newblock The great time series classification bake off: a review and experimental evaluation of recent algorithmic advances. 
\newblock \emph{Data Mining and Knowledge Discovery}, 31(3):606--660, 2017.

\bibitem{Bhattacharya2012}
A.~Bhattacharya and R.~Bhattacharya.
\newblock \emph{Nonparametric Inference on Manifolds with Applications to Shape Spaces}. 
\newblock Cambridge University Press, 2012.

\bibitem{Bellman1962}
\newblock R.~Bellman. 
\newblock On the approximation of curves by line segments using dynamic programming.
\newblock \emph{Communications of the ACM}, 4(6):284, 1961.

\bibitem{Brill2019}
M.~Brill, T.~Fluschnik, V.~Froese, B.~Jain, R.~Niedermeier, and D.~Schultz.
\newblock Exact Mean Computation in Dynamic Time Warping Spaces.
\newblock \emph{Data Mining and Knowledge Discovery}, 33(1):252-291, 2019.

\bibitem{Bulteau2018}
L.~Bulteau, V.~Froese, and R.~Niedermeier.
\newblock Hardness of Consensus Problems for Circular Strings and Time Series Averaging.
\newblock \emph{CoRR}, abs/1804.02854, 2018.

\bibitem{Chakrabarti2002}
K.~Chakrabarti, E.~Keogh, S.~Mehrotra, and M.~Pazzani.
\newblock Locally Adaptive Dimensionality Reduction for Indexing Large Time Series Databases. 
\newblock \emph{ACM Transactions on Database Systems}, 27(2):188--228, 2002.

\bibitem{Chen2015}
Y.~Chen, E.~Keogh, B.~Hu,  N.~Begum, A.~Bagnall, A.~Mueen, and G.E.~Batista. 
\newblock \emph{The UCR Time Series Classification Archive}. 
\newblock \url{www.cs.ucr.edu/~eamonn/time_series_data/}, Accessed: 08/2018.

\bibitem{Cuturi2017}
M.~Cuturi and M.~Blondel.
\newblock Soft-DTW: A Differentiable Loss Function for Time-Series.
\newblock \emph{Proceedings of the 34th International Conference on Machine Learning}, 70:894--903, 2017.

\bibitem{Dau2019}
H.A.~Dau, E.~Keogh, K.~Kamgar, C.-C.M.~Yeh, Y.~Zhu, S.~Gharghabi , C.A.~Ratanamahatana, Y.~Chen, B.~Hu, N.~Begum, A.~Bagnall , A.~Mueen, G.~Batista, and Hexagon-ML. 
\newblock The UCR Time Series Classification Archive. 
\newblock \url{https://www.cs.ucr.edu/~eamonn/time_series_data_2018/}

\bibitem{Dryden1998}
I.L.~Dryden and K.V.~Mardia. 
\newblock \emph{Statistical shape analysis}, Wiley, 1998.

\bibitem{Feragen2013}
A.~Feragen, P.~Lo, M.~De Bruijne, M.~Nielsen, and F.~Lauze.
\newblock Toward a theory of statistical tree-shape analysis.
\newblock \emph{IEEE Transaction of Pattern Analysis and Machine Intelligence}, 35:2008--2021, 2013.

\bibitem{Ferrer2010}
M.~Ferrer, E.~Valveny, F.~Serratosa, K.~Riesen, and H.~Bunke.
\newblock Generalized median graph computation by means of graph embedding in vector spaces. 
\newblock \emph{Pattern Recognition}, 43(4):1642--1655, 2010.

\bibitem{Frechet1948}
M.~Fr\'{e}chet.
\newblock Les \'el\'ements al\'eatoires de nature quelconque dans un espace distanci\'e.
\newblock \emph{Annales de l'institut Henri Poincar\'e}, 215--310, 1948.

\bibitem{Hautamaki2008}
V.~Hautamaki, P.~Nykanen, P.~Franti.
\newblock Time-series clustering by approximate prototypes.
\newblock \emph{International Conference on Pattern Recognition}, 1--4, 2008.

\bibitem{Huckemann2010}
S.~Huckemann, T.~Hotz, and A.~Munk.
\newblock Intrinsic shape analysis: Geodesic PCA for Riemannian manifolds modulo isometric Lie group actions.
\newblock \emph{Statistica Sinica}, 20:1--100, 2010.

\bibitem{Jain2016a}
B.~Jain.
\newblock Statistical graph space analysis.
\newblock \emph{Pattern Recognition} 60:802--812, 2016.

\bibitem{Jain2018}
B.~Jain and D.~Schultz.
\newblock Asymmetric learning vector quantization for efficient nearest neighbor classification in dynamic time warping spaces.
\newblock \emph{Pattern Recognition}, 76:349--366, 2018.

\bibitem{Jain2019}
B.~Jain.
\newblock Revisiting inaccuracies of time series averaging under dynamic time warping.
\newblock \emph{Pattern Recognition Letters}, 125, 418--424, 2019.

\bibitem{Jain2020}
B.~Jain and D.~Schultz.
\newblock Sufficient conditions for the existence of a sample mean of time series under dynamic time warping.
\newblock \emph{Annals of Mathematics and Artificial Intelligence}, 1--34, 2020.

\bibitem{Jiang2001}
X.~Jiang, A.~Munger, and H.~Bunke. 
\newblock On median graphs: properties, algorithms, and applications. 
\newblock \emph{IEEE Transactions on Pattern Analysis and Machine Intelligence}, 23(10):1144-1151, 2001.

\bibitem{Kendall1984}
D.G.~Kendall. 
\newblock Shape manifolds, procrustean metrics, and complex projective spaces. 
\newblock \emph{Bulletin of the London Mathematical Society}, 16:81--121, 1984.

\bibitem{Keogh2001}
E.~Keogh, K.~Chakrabarti, S.~Mehrotra, M.~Pazzani, and S.~Mehrotra.
\newblock Dimensionality Reduction for Fast Similarity Search in Large Time Series Databases. 
\newblock \emph{Journal of Knowledge and Information Systems}, 3(3):263--286, 2001.

\bibitem{Kim2010}
P.T.~Kim and J.-Y.~Koo. 
\newblock Comment on \cite{Huckemann2010}.
\newblock \emph{Statistica Sinica}, 20:72--76, 2010.

\bibitem{Kingma2015}
D. P.~Kingma and  J. L.~Ba.
\newblock Adam: a Method for Stochastic Optimization. 
\newblock \emph{International Conference on Learning Representations}, 2015.

\bibitem{Liu2019}
Y.~Liu, Y.~Zhang, and M.~Zeng. 
\newblock Adaptive Global Time Sequence Averaging Method Using Dynamic Time Warping.
\newblock \emph{IEEE Transactions on Signal Processing}, 67(8):2129--2142, 2019.

\bibitem{Marron2014}
J.S.~Marron, A.M.~Alonso.
\newblock Overview of object oriented data analysis.
\newblock \emph{Biometrical Journal}, 56(5):732--753, 2014.

\bibitem{Wang2007}
H.~Wang and J.S.~Marron.
\newblock Object oriented data analysis: sets of trees. 
\newblock \emph{The Annals of Statistics} 35:1849--1873, 2007.

\bibitem{Petitjean2011}
F.~Petitjean, A.~Ketterlin, and P.~Gancarski. 
\newblock A global averaging method for dynamic time warping, with applications to clustering.
\newblock \emph{Pattern Recognition} 44(3):678--693, 2011.

\bibitem{Petitjean2016}
F.~Petitjean, G.~Forestier, G.I.~Webb, A.E.~Nicholson, Y.~Chen, and E.~Keogh.
\newblock Faster and more accurate classification of time series by exploiting a novel dynamic time warping averaging algorithm. 
\newblock \emph{Knowledge and Information Systems}, 47(1):1--26, 2016.

\bibitem{Rabiner1979}
L.R.~Rabiner and J.G. Wilpon.
\newblock Considerations in applying clustering techniques to speaker-independent word recognition. 
\newblock \emph{The Journal of the Acoustical Society of America}, 66(3):663--673, 1979.

\bibitem{Rebagliati2012}
N.~Rebagliati, A.~Sol\'e, M.~Pelillo, and F.~Serratosa. 
\newblock On The Relation Between The Common Labelling and The Median Graph.
\newblock \emph{Structural, Syntactic, and Statistical Pattern Recognition}, 2012.

\bibitem{Sakoe1978}
H.~Sakoe and S.~Chiba. 
\newblock Dynamic programming algorithm optimization for spoken word recognition. 
\newblock \emph{IEEE Transactions on Acoustics, Speech, and Signal Processing}, 26(1):43--49, 1978.

\bibitem{Schultz2018}
D.~Schultz and B.~Jain.
\newblock Nonsmooth analysis and subgradient methods for averaging in dynamic time warping spaces.
\newblock \emph{Pattern Recognition}, 74:340--358, 2018.
 
\bibitem{Soheily-Khah2016}
S.~Soheily-Khah, A.~Douzal-Chouakria, and E.~Gaussier.
\newblock Generalized $k$-means-based clustering for temporal data under weighted and kernel time warp.
\newblock \emph{Pattern Recognition Letters}, 75:63--69, 2016.

\bibitem{Tan2017}
C. W.~Tan, G.I.~Webb, and F.~Petitjean. 
\newblock Indexing and classifying gigabytes of time series under time warping. 
\newblock \emph{Proceedings of the 2017 SIAM International Conference on Data Mining}, 282--290, 2017.

\bibitem{Terzi2006}
\newblock E.~Terzi and P.~Tsaparas. 
\newblock Efficient algorithms for sequence segmentation. 
\newblock \emph{Proceedings of the 2006 SIAM International Conference on Data Mining}, 316--327, 2006.

\bibitem{Vintsyuk1968}
T.K.~Vintsyuk. 
\newblock Speech discrimination by dynamic programming. 
\newblock \emph{Cybernetics}, 4(1):52--57, 1968.

\bibitem{Wang2011}
Haizhou~Wang and Mingzhou~Song.
\newblock Ckmeans.1d.dp: Optimal $k$-means Clustering in One Dimension by Dynamic Programming.
\newblock \emph{The R Journal}, 3(2):29--33, 2011.

\bibitem{Wilpon1985}
J.G.~Wilpon and L.R.~Rabiner.
\newblock A Modified $K$-Means Clustering Algorithm for Use in Isolated Work Recognition.
\newblock \emph{IEEE Transactions on Acoustics, Speech, and Signal Processing}, 33(3):587--594, 1985.

\bibitem{Yi2000}
B.-Y.~Yi and C.~Faloutsos. 
\newblock Fast time sequence indexing for arbitrary $L_p$ norms. 
\newblock \emph{International Conference on Very Large Databases}, 385--394, 2000.
\end{thebibliography}
\end{document}